\documentclass{aa} 
\usepackage{graphicx}
\usepackage{rotating,amssymb,calc}
\usepackage{natbib} 
\bibpunct{(}{)}{;}{a}{}{,} 
\begin{document}
   \title{High-redshift quasar host galaxies with adaptive optics}
   
   \author{B. Kuhlbrodt\inst{1,2}
          \and
          E. \"Orndahl\inst{3,4}
          \and
          L. Wisotzki\inst{1,5}
          \and
          K. Jahnke\inst{1,2}
          }

   \offprints{E. \"Orndahl,
   \email{evaorn@utu.fi}}

   \institute{Astrophysikalisches Institut Potsdam, An der Sternwarte 16, 
        14482 Potsdam, Germany
   \and Hamburger Sternwarte, Gojenbergsweg 112, 21029 Hamburg, Germany
   \and   Tuorla Observatory,  University of Turku, 
           V\"ais\"al\"antie 20, FI-21500 Piikki\"o, Finland
   \and Department of Astronomy and Space Physics, Uppsala University,
              Box 515, 751 20 Uppsala, Sweden
      \and Institut f\"ur Physik, Potsdam University, Am Neuen Palais 10, 
        14469 Potsdam, Germany 
           }

   \date{Received sometime; accepted later}

   \abstract{ We present $K$ band adaptive optics observations of three 
     high-redshift ($z \sim 2.2$) high-luminosity quasars, all of which 
     were studied for the first time. We also observed several 
     point spread function (PSF) calibrators, non-simultaneously because 
     of the small field of view. The significant temporal PSF variations 
     on timescales of minutes inhibited a straightforward scaled PSF removal 
     from the quasar images. Characterising the degree of PSF concentration 
     by the radii encircling 20~\% and 80~\% of the total flux, respectively,
     we found that even under very different observing conditions  
     the $r_{20}$ vs.\ $r_{80}$ relation varied coherently between 
     individual short exposure images,
     delineating a well-defined relation for point sources. 
     Placing the quasar images on this relation, we see indications
     that all three objects were resolved. We designed a procedure
     to estimate the significance of this result, and to estimate
     host galaxy parameters, by reproducing the statistical distribution
     of the individual short exposure images.
     We find in all three cases evidence for a luminous host galaxy, 
     with a mean absolute magnitude of $M_R = -27.0$ and scale lengths
     around $\sim 4$--12~kpc. Together with a rough estimate of the
     central black hole masses obtained from C$\:${\sc iv} line widths,
     the location of the objects on the bulge luminosity vs.\ black hole 
     mass relation is not significantly different from the low-redshift 
     regime, assuming only passive evolution of the host galaxy. 
     Corresponding Eddington luminosities are 
     $L_\mathrm{nuc}/L_\mathrm{Edd} \sim 0.1$--0.6.

   }

   \maketitle
%

\section{Introduction}

The study of high-redshift quasar host galaxies has during the past
few years become quite an active research field, offering new roads of
insight both on the phenomenon of quasar evolution as well as galaxy
formation in the early universe.  The strong cosmic evolution seen in
the quasar population from the peak at $z\sim$ 2--3 is very likely an
effect of changing environmental conditions. Together with the
evolution in the star formation rate from $z\sim$~2 to the present,
this implies a strong link between the formation and subsequent
evolution of galaxies and the processes that trigger and maintain the
quasar activity \citep[e.g.][]{fran99}. This is reflected in the
correlation between black hole mass and host spheroid luminosity found
for local massive ellipticals \citep{mago98,korm01}, which has been shown 
to hold for also quasars \citep{laor98,mclu02}. Observations of quasar 
host galaxies over a range of redshifts are necessary to obtain further
insights into these links.

It has been repeatedly predicted that the observational technique of 
adaptive optics (AO) could be pivotal in the advancement of quasar
host galaxy research, because of its greatly enhanced angular resolution.
AO observations are already successfully employed for the imaging of,
e.g., protoplanetary disks and binary star systems, but it is probably
fair to say that there are still only very few applications to quasar hosts.
One reason for this could well be due to an inherent property of
AO images of point sources, namely the superposition of a 
diffraction-limited core with an extended halo. 
Operating at low Strehl ratio (often unavoidable at high Galactic latitudes
where bright guide stars for wavefront sensing are scarce) greatly 
enhances this problem.  Furthermore, accurate
knowledge of the point spread function (PSF) is a prerequisite.
However, the high spatial sampling of AO detectors is always paid for
with a small field of view, and non-simultaneous PSF observations
often cannot be avoided. As a consequence, the fundamental task 
of correctly differentiating between a compact nucleus and an 
extended host galaxy remains a challenge.

Most of the existing AO data on quasar hosts was obtained in the
near-infrared $K$ band, where the effects of atmospheric turbulence
are less pronounced than at shorter wavelengths.  In the case of low
to intermediate redshift quasars, the host galaxies are easily
resolved, and the PSF details were not crucial to the detection of the
objects \citep{stoc98,marq01}; rather, the main emphasis was on the
detection of structural details. This is different at high redshifts
($z \ga 1$) where the detection of the host galaxy as such becomes
again a challenge, even with AO \citep{aret98b,hutc01,lacy02,croo04,falo05}.  
Other ground- and space-based campaigns with high 
resolution were also successful in resolving host galaxies to $z \sim 2.5$
\citep{kuku01,jahn04,falo04}, or even $z \ga 4$ \citep{hutc03}.

The focus of the present paper is on high-luminosity quasars 
around and slightly above $z\sim 2$, with the aim of constraining 
the host vs.\ nuclear luminosity (and black hole mass) relations 
at these redshifts. Although we used an AO instrument/telescope 
combination that has been decommissioned since, 
published AO observations of high-$z$ quasars are still so scarce 
that even our small sample makes a non-negligible contribution
to the field. Moreover, we have spent considerable effort at
analysing the effects of temporal PSF variability, and designed
a method to infer on quasar host properties with non-simultaneously
obtained PSF observations. Besides our astrophysical results, 
we hope that some of the methodical insights and strategic aspects
mentioned in this paper may be of interest to the community.

We first present our set of new targets in Sect.~\ref{sec:obs}, together 
with a summary of the AO observations and data reduction.
Section~\ref{sec:PSF} is dedicated to the task of image quality
assessment, emphasising the aspects of temporal PSF variability.
In Sect.~\ref{sec:sim} we describe our analysis strategy, based
on extensive simulations designed to reproduce the observed image
characteristics. The results of applying this method to our data
are presented in Sect.~\ref{sec:quasars}, which we discuss in 
astrophysical context in Sect.~\ref{sec:disc}. Throughout
this paper we adopt $H_0= 70 $~km~s$^{- 1}$~Mpc$^{-1}$, $\Omega_m=0.3$
and $\Omega_\Lambda=0.7$. All magnitudes are zeropointed to the Vega system.

\section{Targets}  \label{sec:obs}

\begin{figure}
\centerline{%
\includegraphics*[width=88mm]{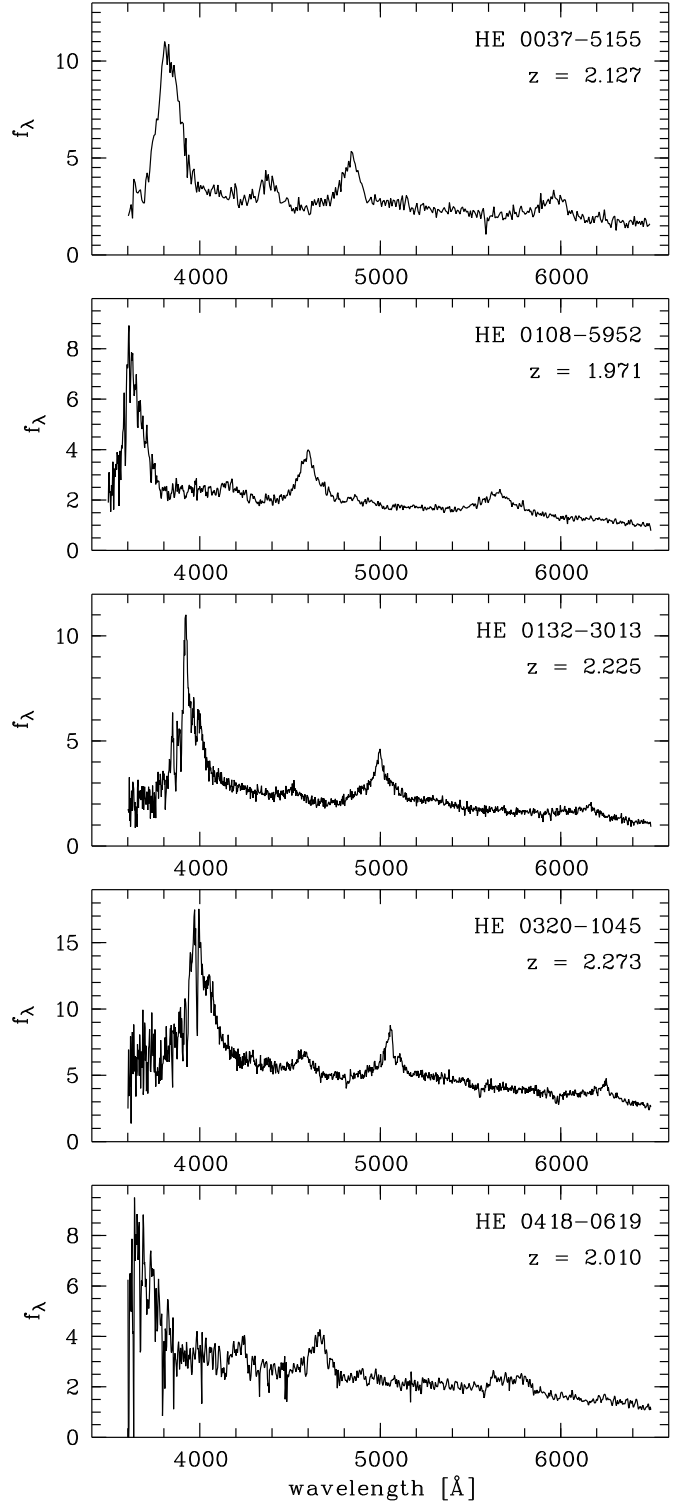}}
\caption[]{Slit spectra of the five newly discovered quasars, taken
with ESO telescopes. See text and Table~\ref{tab:sample} for
details. $f_\lambda$ is in units of $10^{-16}$~erg~cm$^{-2}$~s$^{-1}$.
\label{fig:spectra}}
\end{figure}

\begin{figure*}
\setlength{\unitlength}{1mm}
\begin{picture}(180,38)
\put(0,0){\includegraphics*[width=35mm]{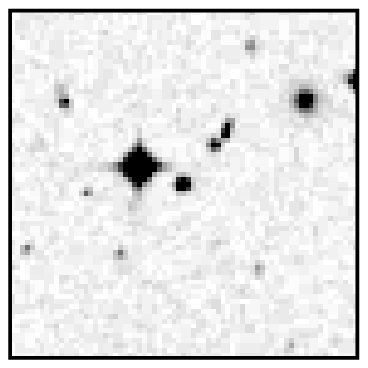}}
\put(6,36){HE 0037$-$5155}
\put(36,0){\includegraphics*[width=35mm]{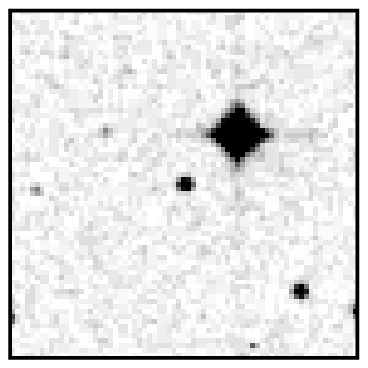}}
\put(42,36){HE 0108$-$5952}
\put(72,0){\includegraphics*[width=35mm]{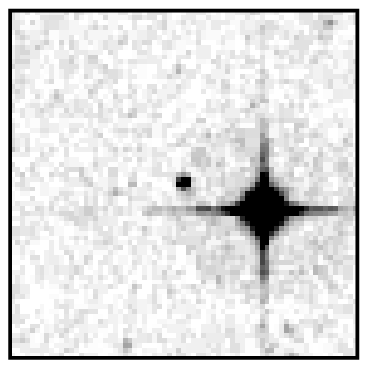}}
\put(78,36){HE 0132$-$3013}
\put(108,0){\includegraphics*[width=35mm]{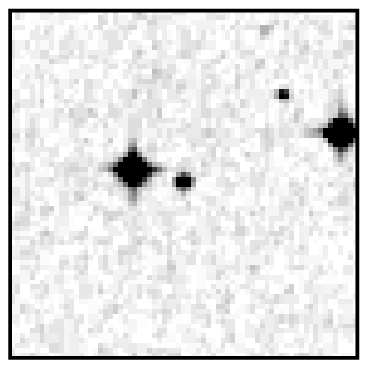}}
\put(114,36){HE 0320$-$1045}
\put(144,0){\includegraphics*[width=35mm]{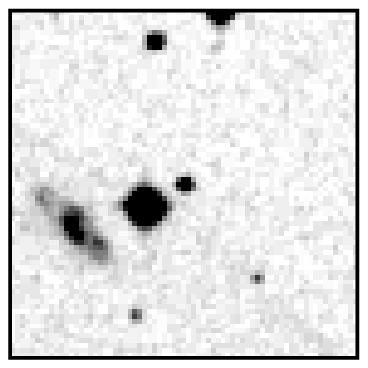}}
\put(150,36){HE 0418$-$0619}
\end{picture}
\caption[]{$2'\times 2'$ DSS images around the five featured quasars
(the central object in each panel), together with their corresponding 
AO guide stars. North is up, East is to the left.
\label{fig:dss}}
\end{figure*}

\setlength{\tabcolsep}{3mm}
\begin{table*}
\centering
    \caption{ Basic properties of the input target sample quasars, PSF stars, and the corresponding
      adaptive optics guide stars. Column 5 gives $B_J$ band magnitude for
      the quasars and $V$ band magnitude for the PSF stars. The last
    three PSF stars were observed unrelated to any of the quasars.
      }
    \begin{tabular}{lccccccc}
\noalign{\smallskip}\hline
\noalign{\smallskip}\hline\noalign{\smallskip}
      Object name    & \multicolumn{2}{c}{\hfill R.A.\hfill(J2000.0)\hfill Dec\hfill~} &
                        $z$ & $B_J$/$V_\star$ & Guide star & distance & $V_\star$ \\
\noalign{\smallskip}\hline\noalign{\smallskip}
 HE 0037$-$5155 & $00^\mathrm{h}\:40^\mathrm{m}\:17\fs 1$ & $-51\degr\:38'\:49''$ & 
                         2.127 & 17.7 & GSC0803000529 & 16\farcs 8 & 12.9 \\
 GSC0802400921& $00^\mathrm{h}\:37^\mathrm{m}\:34\fs 7$& $-47\degr\:19'\:44''$& 
                               & 14.8 & GSC0802400297 & 15\farcs 2 & 13.0 \\
 GSC0803000534& $00^\mathrm{h}\:41^\mathrm{m}\:21\fs 7$& $-52\degr\:28'\:00''$& 
                               & 15.6 & GSC0803000242 & 13\farcs 2 & 13.8 \\[1ex]
 HE 0108$-$5952 & $01^\mathrm{h}\:10^\mathrm{m}\:52\fs 0$ & $-59\degr\:36'\:21''$ &
                         1.971 & 18.9 & GSC0847901239 & 25\farcs 2 & 12.0 \\[1ex]
 HE 0132$-$3013 & $01^\mathrm{h}\:34^\mathrm{m}\:33\fs 8$ & $-29\degr\:58'\:15''$ &
                         2.229 & 18.0 & GSC0642801994 & 28\farcs 8 & 11.0 \\
 GSC0642602115& $01^\mathrm{h}\:26^\mathrm{m}\:21\fs 8$& $-25\degr\:49'\:46''$& 
                               & 14.6 & GSC0642601594 & 26\farcs 1 & 11.5 \\
 GSC0700200978& $01^\mathrm{h}\:19^\mathrm{m}\:57\fs 9$& $-30\degr\:44'\:50''$& 
                               & 15.5 & GSC0700200865 & 27\farcs 2 & 11.7 \\[1ex]
 HE 0320$-$1045 & $03^\mathrm{h}\:22^\mathrm{m}\:24\fs 5$ & $-10\degr\:35'\:12''$ &
                         2.282 & 17.0 & GSC0529800301 & 18\farcs 0 & 13.0 \\
 GSC0530900226&$03^\mathrm{h}\:39^\mathrm{m}\:24\fs 1$& $-12\degr\:39'\:49''$& 
                               & 14.9 & GSC0530900199 & 15\farcs 6 & 13.5 \\
 GSC0529400760&$03^\mathrm{h}\:05^\mathrm{m}\:37\fs 7$& $-08\degr\:02'\:02''$& 
                               & 14.8 & GSC0529400764 & 19\farcs 9 & 13.2 \\[1ex]
 HE 0418$-$0619 & $04^\mathrm{h}\:21^\mathrm{m}\:24\fs 2$ & $-06\degr\:12'\:04''$ &
                         2.010 & 19.0 & GSC0473301854 & 15\farcs 6 & 11.9 \\
 GSC0472901414&$04^\mathrm{h}\:11^\mathrm{m}\:08\fs 3$& $-03\degr\:46'\:53''$& 
                               & 13.8 & GSC0472901410 & 16\farcs 7 & 13.1 \\[1ex]
 GSC0548301352& $09^\mathrm{h}\:54^\mathrm{m}\:37\fs 1$& $-11\degr\:18'\:32''$& 
                               & 14.3 & GSC0548301353 & 26\farcs 2 & 11.6 \\
 GSC0490901021& $10^\mathrm{h}\:00^\mathrm{m}\:44\fs 5$& $-07\degr\:01'\:05''$& 
                               & 14.6 & GSC0490901030 & 26\farcs 6 & 11.8 \\
 GSC0490900476& $10^\mathrm{h}\:04^\mathrm{m}\:58\fs 0$& $-05\degr\:17'\:44''$& 
                               & 14.9 & GSC0490900457 & 27\farcs 9 & 11.7 \\

\noalign{\smallskip}\hline
    \end{tabular}
    \label{tab:sample}
\end{table*}
\setlength{\tabcolsep}{2.5mm}

\subsection{Target selection}

The number of high-redshift quasars bright enough to allow on-axis
adaptive wavefront correction is extremely small. It is basically zero for
the more venerable AO systems such as ADONIS on the ESO 3.6~m
telescope used by us (see below).  In line with other researchers, we
have therefore searched for targets with a nearby bright star that can
be used for wavefront sensing. For ADONIS, the uttermost limits for AO
correction in the $K$ band were: a red ($R$ or $I$) stellar magnitude
$\la 13$, and a distance between quasar and star of $\la 30''$; to be
practical, these two criteria should not be carried to their extremes
together. Even then, the low surface densities of pairs of quasars and
stars fulfilling both conditions make the number of such targets
necessarily small in any survey. Other groups before us already
used the available quasar catalogues to identify possible targets, and
at the time of planning this project in 1999, these catalogues seemed
to be more or less exhausted, at least to the limits of systems such
as ADONIS.

We therefore performed a new search using the largely unexplored
database of optically bright quasars from the Hamburg/ESO survey
\citep[HES,][]{wiso00},
containing several thousand bright quasars and
quasar candidates at all redshifts up to $z\simeq 3.2$. The survey
magnitude limit of $B_J \la 18$ (on average) makes it a rich source of
high-luminosity quasars. We went through the full database, selecting
all quasars with redshifts $z\ga 1.8$, and included also quasar
candidates where a tentative redshift had been assigned based on their
digitised objective prism spectra \citep[for details,
see][]{wiso96,wiso00}.  We then paired the list with the \emph{HST
Guide Star Catalogue} (GSC) and searched for pairs matching the above
given criteria for possible ADONIS guide stars with $B_J < 14$.
Altogether, we arrived at 12 candidate targets selected by these
criteria. Of these, seven objects strained both the magnitude and the
distance limits to an unacceptable degree and were eliminated from the
list. We were then left with five remaining targets that we eventually
took to the telescope.

All of these five quasars appear here for the first time in the
literature.  The data for them are given here for the benefits of the
community, as we believe that they might be
interesting to other groups working in the field of quasar host galaxies
and adaptive optics in the future. Table~\ref{tab:sample} contains
their basic properties, including those of the adopted AO guide star.
The quasars are drawn from an optical survey, 
and no radio flux measurements of the sources exist; they
are most likely radio-quiet.
We show slit spectra of the quasars in Fig.\ \ref{fig:spectra}, taken
with either the ESO 1.52~m or the ESO/Danish 1.54~m telescope between
1996 and 2000. Further below in this paper we use these spectra to
obtain a rough estimate of the black hole masses in these objects.
Finally, Fig.\ \ref{fig:dss} shows postage stamp images of the 
quasars together with their nearby AO guide stars.

\subsection{PSF calibrators}\label{sec:psfcali}

In order to obtain photometry and morphological parameters of a quasar
host galaxy, the contribution of the point-like active nucleus has to
be removed, for which an accurate knowledge of the PSF is essential.
In our case with a field of view of only $12\farcs8\times
12\farcs8$, the quasars were in all cases completely isolated,
and we were forced to obtain constraints of the PSF from 
external stellar calibrators, to be observed non-simultaneously 
with the quasars. (Even if there had been useful stars inside
the field of view, the inevitable spatial PSF variations over 
the isoplanatic patch would have made their use highly questionable.)

We decided to accompany each quasar with two different PSF
calibrators, to enable cross-validation.  Each of these PSF calibrator
stars were selected to have a wavefront sensing guide star of
magnitude and distance as closely matching to that of the quasar guide
star as possible  (see Table~\ref{tab:sample}).
The PSF stars themselves were chosen to be substantially brighter than
the quasars, typically around $B_J \simeq$ 14--15, allowing for a high
S/N PSF definition with a small number of exposures. Following these
criteria, we selected several stellar pairs from a $\sim 5\degr$
surrounding of each quasar.

\subsection{Observations}

Observations were performed in the $K_s$ band on the nights of 1999
November 27--29, using the ADONIS system on the ESO La Silla 3.6m
telescope. The SHARPII+ camera was equipped with a 256$\times$256
Nicmos III array with a pixel size of 40 $\mu$m. The pixel scale was
set to 0\farcs05/pixel, resulting in a field of view of
$12\farcs8\times 12\farcs8$. Because of variable weather conditions
we concentrated on only three of our targets: HE~0037$-$5155, 
HE~0132$-$3013, and HE~0320$-$1045.

For every target the quasar itself was first observed in a cycle of
$10\times60$ or $15\times60$ seconds, followed by the corresponding
two PSF stars in a cycle of $3\times60$ seconds per object. 
Each individual frame of 60~s exposure (which in turn consisted
of ten 6~s detector integration cycles)
was saved and used in the later analysis.
Photometric standard stars were obtained at the beginning and end of 
each night, and skyflats were acquired in evening twilight.  
The seeing as computed from the wavefront sensing data by the 
ADONIS instrument software was stable at $\simeq0\farcs6$ 
during the first night, but varied between
0\farcs7--1\farcs1 and 0\farcs8--1\farcs2, respectively, 
during the useful portions of the following two nights.

To monitor the variable sky background, we used the standard option
of switching back and forth between two locations of the quasar on 
the chip $\sim$8\arcsec\ apart, using the internal chopping mirror. 
Total integration times amounted to 160~min for HE~0037$-$5155, 
170~min for HE~0132$-$3013, and 230~min for HE~0320$-$1045, while
each of the PSF stars was typically integrated for 36~min total.

\subsection{Reduction}

\begin{figure}
\centering
\includegraphics[bb=34 125 576 668,clip,width=8cm]{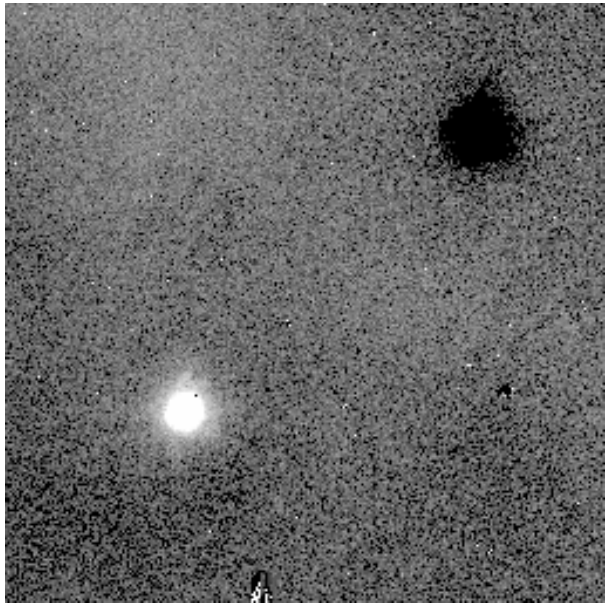}
\caption[]{Example of a background corrected image from an image cube
of HE~0037$-$5155. The quasar is located in the lower left
corner, while the opposite corner contains a negative residual 
from the alternate chopping position.
\label{fig:image}}
\end{figure}

After bias subtraction and flatfielding, each stack of consecutive
1~min integrations of a single target was recorded in two image cubes,
separating odd- and even-numbered frames.
Each frame set was averaged separately.
The odd-numbered average was then adopted as sky background estimate
for the stack of even-numbered images and subtracted, and vice versa.
Since each average frame contained also the quasar image in the opposite
quadrant, the subtraction creates a negative imprint of the quasar 
in all individual frames (Fig.~\ref{fig:image}). This was not a problem
since we used only the quadrant with the quasar in it for further
analysis.

After subtraction of the background frame, some lower level residual
structure remained that was created by different thermal emission 
patterns from the two positions of the chopping mirror \citep[see][]{shar97}. 
Since only a small detector area covered by the quasar light
distribution with a diameter of $\sim 2''$ was needed for analysis, 
we assumed the background to be locally constant. 
The level of this local background was then fine-tuned by demanding
a constant radial curve of growth between 1\arcsec and $1\farcs5$ radius. 
This is a conservative procedure; some small fraction of quasar flux
was clearly still present within this range, and we are thus
slightly biased against the detection of extended emission.
Using simulations, we estimated this effect to be less than 0.1~mag 
for all realistic host galaxy models. Increasing this radius
would have decreased the systematic error, but at the same time
it would have enhanced the vulnerability to small-scale residual
sky variations on an amplitude comparable to the local object flux.

For all reduced frames we created bad pixel maps from skyflats using
the `flat' task from the ECLIPSE data reduction package
\citep{ecli01}. It was found that bad pixel areas changed in size
within image cubes, so a global bad pixel map was therefore
complemented with individual maps for each frame where also cosmics
were marked. The remanence effect common for Nicmos III arrays (which
alters the sensitivity of pixels over- or underexposed in the previous
frame) was found to be negligible for both quasar and PSF star images.
Photometric calibration was
performed by aperture photometry on the standard stars. The
uncertainty of the calibration is 0.05 magnitudes.  Finally, 
the quadrant containing the quasar was extracted from each reduced 
image cube.
During data reduction we found that some single images or whole 
image cubes (of both quasars and stars) were badly affected by problems 
with the AO optimisation loop or with the guiding. These data 
were excluded from subsequent analysis.

\begin{figure}
\includegraphics[clip,width=\columnwidth*9/10]{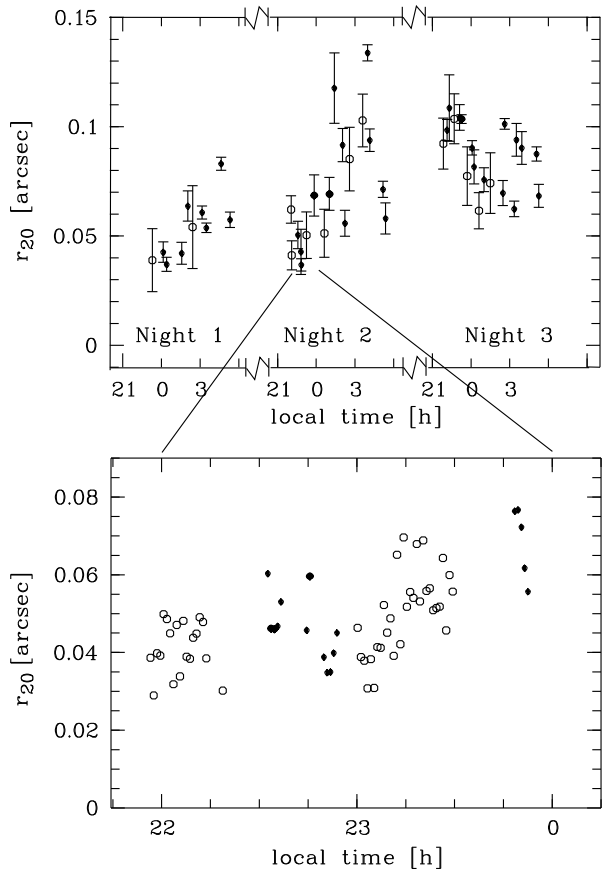}
\caption[]{Progression of core width with time. In the top panel we
  plot average values and $1\sigma$ error bars for each image cube for
  both PSF calibrator stars (filled diamonds) and quasar observations
  (circles).  We expand the beginning of the second night in the lower
  panel, plotting values for every image of the first five cubes. The
  width variation between two consecutive images can be as large as
  30\,\%.
\label{fig:timer20}}
\end{figure}


\section{The point spread function}   \label{sec:PSF}

\subsection{Image quality assessment and PSF variability}

As laid out in Sect.~\ref{sec:psfcali} above, we had to rely
on PSF calibrator stars observed non-simultaneously, but 
with configurations very similar to the quasar observations.
A major concern with this approach lies in the fact that
under permanently changing ambient conditions, the PSF 
itself is expected to vary with time. 
In order to monitor and assess the temporal 
variability of the PSF, the observations of PSF calibrator stars 
and quasars were nested. 

Because of the nearly diffraction-limited core of AO images,
the full-width at half maximum (FWHM) is not a very 
sensitive image quality indicator. A better quantity is
the Strehl ratio $S$, which however is difficult to determine
for quasar images because of the possible contribution of the host
galaxy. We adopted as `core width' indicator the radius $r_{20}$
encircling 20\,\% of the total flux of the object.
For point sources, we found that $r_{20}$ is closely related to 
the Strehl ratio, following the approximate relation  
$S \approx 0.52 - 0.32\: (r_{20}/1'') + 0.056\: (r_{20}/1'')^2$.  
The variation of $r_{20}$ is plotted in 
Fig.~\ref{fig:timer20}. Typical core widths are near 0\farcs05
or slightly above, corresponding to Strehl ratios of around 0.2 
(ranging from below 0.1 to above 0.3). 

 Unfortunately, we found the PSF to be significantly variable even 
within single image cubes, over lapses of 30~min or less. 
%
The reason for the variability of the PSF is the rapid change of the
atmospheric turbulence characteristics, inducing different responses
from the AO system. This in turn leads to variation in the centroid,
higher FWHM, lower Strehl ratios and higher speckle noise.  An
increase in integration time, or coaddition of several images, would
in general reduce this variation but will at the same time decrease
resolution and hence the benefits of AO observation \citep[cf.\ 
][]{mign98}. In the $K_s$ band, exposure times are in any case limited
by the strong sky background emission in order to avoid non-linearity
or saturation effects.  Since the variations also apply to PSF star
images displaying similar Strehl values, thus
suggesting very similar correction quality of the AO system, we have
to conclude that the PSF in any individual quasar image is essentially
unknown, and can only be estimated by statistical means.

This posed a serious problem, as standard methods of quasar host analysis 
can only be applied to AO data when the hosts are very extended and 
the knowledge of the exact shape of the PSF is less important. 
While this is usually the case for low-$z$ quasar
when structures outside of the centre are to be resolved, 
the situation is different for high-$z$ quasars. 
It is clear that another approach was needed to
evaluate whether an observed quasar is extended or not, and to
determine the characteristics of the possible host galaxies.

In order to describe the image PSF we first investigated the
most prominent variation: the variation of image concentration.

\subsection{Image concentration diagnostics}

\begin{figure}
\centering 
\includegraphics[bb=247 428 485 599,clip,width=\columnwidth]{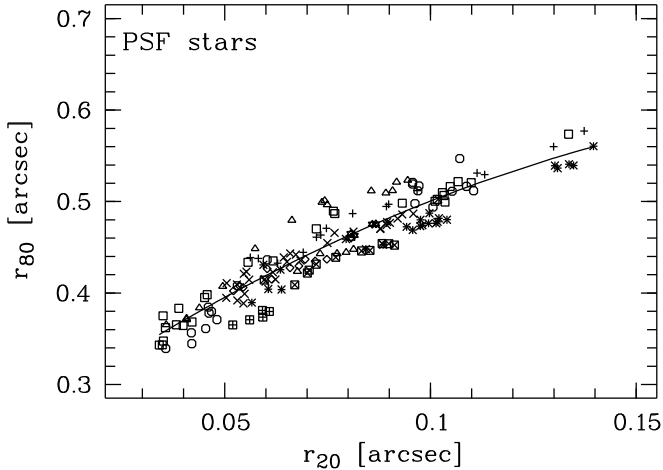}
\caption[]{
The $r_{20}$--$r_{80}$ diagram for the PSF stars. Each point
corresponds to a single 1~min image from an image cube. Different
individual PSF stars are coded with different symbols. The solid line
is the best fit to the PSF star points (`PSF track'). The
variation due to different observing conditions is generally larger than
the difference between individual PSF stars.
\label{fig:r20r80star}}
\end{figure}

While the PSF variation in general is probably not predictable, 
we considered the question whether at least some simplified shape
descriptor might show some regularity. Figure~\ref{fig:r20r80star} 
displays the relation between the width of the core and the width 
of the wings, as measured by the radii including 20~\% and 
80~\% ($r_{80}$) of the flux, for all individual 1~min PSF star
observations. The data clearly follow a -- slightly non-linear -- 
trend, which we approximated with a parametrised polynomial fit in
Figure~\ref{fig:r20r80star}. This fit will be called the
`PSF track' henceforth. Notice that we use both values $r_{20}$ and
$r_{80}$ rather than just their ratio as a `concentration index'
to allow for more freedom in describing the observed shape variations.

Remarkably, Fig.~\ref{fig:r20r80star} contains data taken under very
different ambient conditions, with guide stars of different
brightnesses and with different AO optimisation settings;
nevertheless, the relation is rather tight.  The main reasons for the
scatter of individual frames around this are photon shot noise and
variable two-dimensional asymmetries in the PSF shape, in proportions
depending on the brightness of the object. The PSF star--guidestar
%
%
configuration has much less influence on the position of the
observations in the $r_{20}$--$r_{80}$ diagram than the observing
condition. This can well be seen in PSF stars which were observed in
several nights, e.g.\ GSC0642602115 (coded by triangles) which is placed
both above and below the PSF track.

\begin{figure}
\centering
\includegraphics[bb=47 59 286 230,clip,width=\columnwidth]{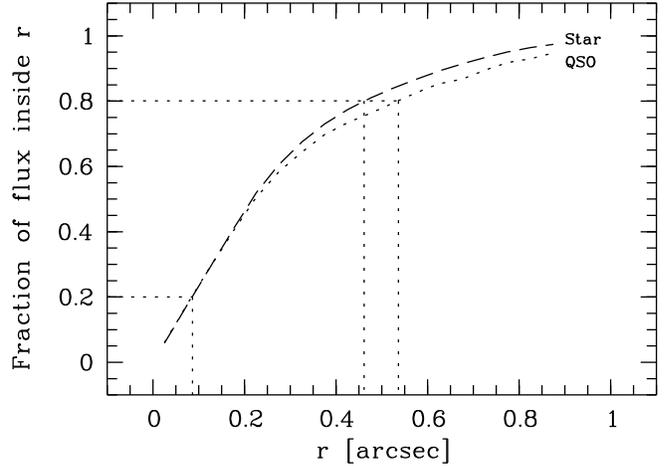}
\caption[]{Growth curves for a star (dashed line) and a quasar (dotted
  line), taken from the first cubes of the second night (lower panel
  in Fig.~\ref{fig:timer20}). The determination of $r_{20}$ and
  $r_{80}$ radii allows a distinction between the two types.
\label{fig:r20r80}}
\end{figure}

Given that the PSF can be characterised by such a simple relation, 
we now demonstrate that the same diagnostic can be used to search for
faint extended emission underlying a bright quasar image.
Figure~\ref{fig:r20r80} shows the curves of growth of both
a PSF star and of a quasar. While both are dominated in their cores
by the PSF, thus having virtually identical $r_{20}$ values, 
the host galaxy contributes fractionally more flux to the wing, 
thus flattening the curve of growth and increasing $r_{80}$. 

\begin{figure*}
\includegraphics[bb=56 574 513 716,clip,width=\textwidth]{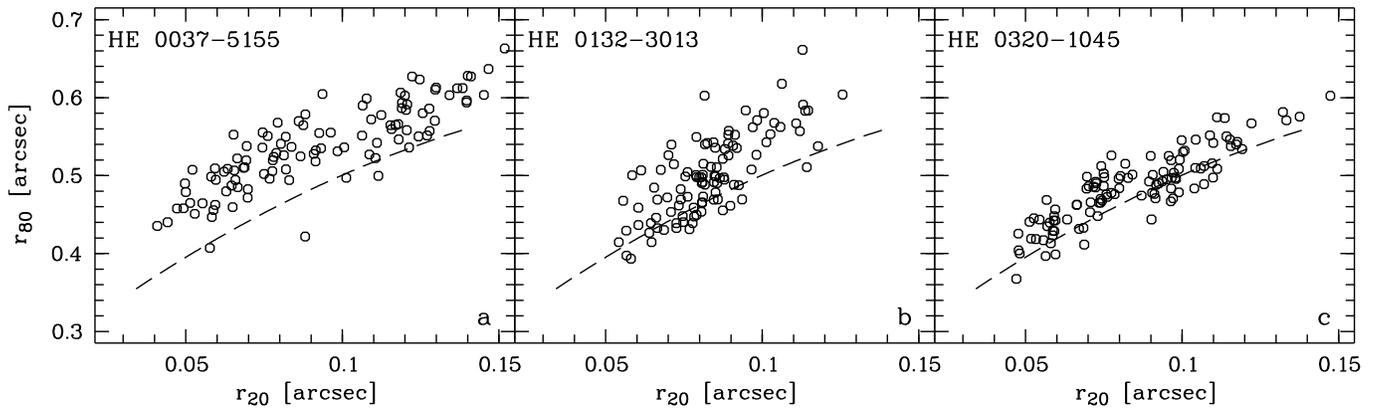}
\caption[]{The $r_{20}$--$r_{80}$ diagrams for the quasars.  Each point corresponds to a single image
  from an image cube. The dashed line is the fitted `PSF track' from
  Fig.~\ref{fig:r20r80star}.
\label{fig:r20r80qso}}
\end{figure*}

In Fig.~\ref{fig:r20r80qso} we plotted $r_{80}$ vs.\ $r_{20}$ 
for all individual 1~min quasar images of the three nights.
We show the distribution of points for the three quasars separately
(Figs.~\ref{fig:r20r80qso}\,a--c), where it can be seen that for none
of them the points scatter \emph{around} the fitted PSF track.
In the cases of HE~0037$-$5155 and HE~0132$-$3013 the offset between
points and line appears significant even at first glance, whereas 
for HE~0320$-$1045 the effect is less pronounced and debatable.
Notice that the quasars show a much higher scatter than
the PSF stars in Fig.~\ref{fig:r20r80star}; this is expected,
since the stars are typically ten times brighter than the quasars. 
This implies that the scatter in the quasar images is dominated
by shot noise.

The shift of the quasars away from the PSF track in these diagrams is
immediately suggestive of influence from a host galaxy, but in order
to quantitatively confirm a detection, we need a better understanding
of the principles which create the distribution of points in the
diagram.  To this end we will in the following attempt to reconstruct
the distribution of quasars in these diagrams using our knowledge
about the distribution of PSF star points under the hypothesis that
the presence of a host galaxy is responsible for the observed shift.
This is done in three steps, by investigating each of the following 
questions in turn:

\begin{itemize}
\item Can the offsets of the quasar images be explained by
  adding host galaxy flux to a point source?
\item Is one galaxy model able to explain the average
  $r_{20}$--$r_{80}$ relation for a given quasar?
\item Can the entire distribution of quasar data points be represented
  with a set of simulations created to match the actual conditions of
  observation?
\end{itemize}

\section{Simulations}
\label{sec:sim}
\subsection{Individual images}

To simulate a single quasar observation, we created artificial quasar
nuclei with surrounding host galaxies as would be observed under
different external conditions, and studied their behaviour in the
$r_{20}$--$r_{80}$ diagram. The artificial objects were composed of the
observed image of a PSF star to represent the nucleus, plus a model
galaxy. The numerical model we use is the well-known spheroidal law by
\citet{deva48}:
\begin{equation}   
 F_{\mathrm{sph}}(r) = F_{\mathrm{sph},0}\,\exp\left[-7.67\,\left(\frac{r}{r_{5
0}}\right)^{1/4}\right],
 \label{eqn:sph}   
\end{equation}   
where $r_{50}$ is the radius which encircles half the total flux. 

We chose the spheroidal model since it is not unreasonable to expect
highly luminous quasars to be hosted by ellipticals
\citep{mcle95, dunl03}. Even if the host galaxies at high redshifts
are more disturbed, the above law can still be considered appropriate 
as a description of the main part of the flux
\citep[e.g.][]{hutc02}, since low surface brightness disk components
or tidal features are likely to be missed anyway. Even in the case of
a true disk-type host we probably could not distinguish it from an 
elliptical of similar size and brightness.

We varied the half-light radius $r_{50}$, but restricted the tests to
circularly symmetric galaxies for simplicity. The models were numerically 
convolved with an empirical PSF given by an arbitrary PSF star image, 
to create a light distribution consistent with the external conditions. 
Star and convolved object were then scaled and added to mimic different 
ratios of nuclear to galaxy flux.

The outcome can be seen in Fig.~\ref{fig:grid}. Each tickmark along
the solid lines marks the position of an individual simulated image in
the $(r_{20},r_{80})$ parameter space. The images along a given solid
line were constructed having the same flux ratio between nucleus and
host galaxy. They differ only in the PSF star image used in their
construction, hence only in external observing conditions. After
creation, the simulated data was processed in a manner identical to
the real data to extract the $r_{20}$ and $r_{80}$ parameters (the
data for HE~0037$-$5155 is overplotted in Fig.~\ref{fig:grid} 
for comparison). The different solid lines correspond to sets of
simulations with different ratios of nuclear to host flux (n/h). 
This simulation shows that the artificial quasars, in principle, 
cover the same region in the $r_{20}$--$r_{80}$ diagram as 
the real objects.

\subsection{Image ensembles}  \label{sec:simavg}

Since neither the nucleus nor the host galaxy vary intrinsically
during the observation time span, there must be \emph{one} galaxy model, 
viewed under different observing conditions, that is sufficient 
to represent the distribution of a quasar observation.

\begin{figure}
\centering
\includegraphics[bb=51 284 285 739,clip,width=\columnwidth]{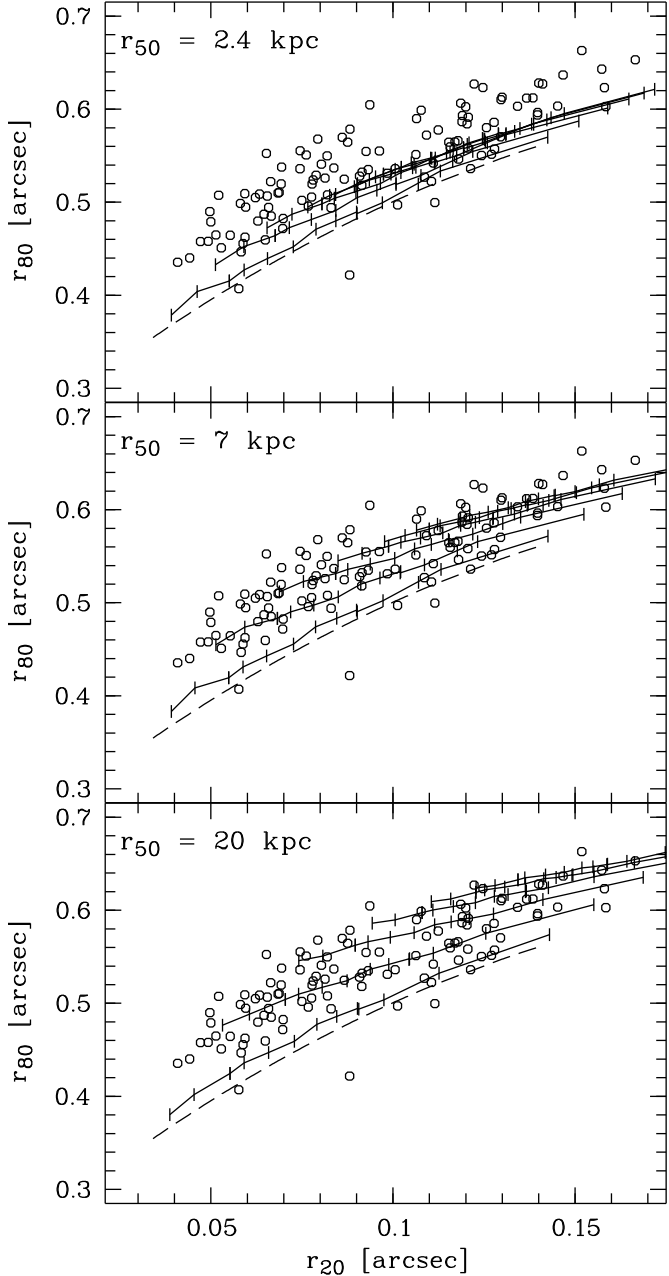}
\caption[]{Loci for artificial objects with three different model galaxy 
  scale lengths, plotted on top of the data for HE\,0037--5155. The
  scale lengths were computed at the redshift of the quasar.  The
  dashed line is the PSF track, and the solid lines represent different
  ratios of nucleus to host flux with the tickmarks showing the
  location of each object. The n/h ratios 
  are 7.7, 1.5, 0.66, 0.38, 0.23 and 0.16, counting from the bottom to the
  top.
  \label{fig:grid}}
\end{figure}

In Fig.~\ref{fig:grid} we show the result for model galaxy half-light
radii of 2.4, 7 and 20~kpc, using fifteen different PSF star images 
to represent the range of observing conditions, and six
different flux ratios n/h. The PSF star images were selected
to be close to the PSF track (we shall relax that condition in the
next subsection) and to be spaced roughly equidistantly along the PSF track. 
The tickmarked lines in the plot show the set of $r_{20}$--$r_{80}$ 
values extracted for each artificial object class. The further away 
a line lies from the PSF track, the lower is the n/h ratio, and thus 
the higher is the host galaxy flux.

With increasing host galaxy flux, the lines of constant n/h
ratio flatten, and they move upwards and to the right in the
$r_{20}$--$r_{80}$ diagram. This flattening can be understood when
considering the locus of the extreme case of a pure galaxy without any
nucleus. In the limiting case of very bad
seeing (high $r_{20}$ values), any galaxy will become unresolved, and
galaxy and star become indistinguishable. Thus the `pure galaxy line'
and the PSF track join. When external conditions improve, the $r_{20}$
values become smaller and the galaxy starts to be resolved.
The difference in $r_{80}$ between point source and galaxy increases,
resulting in a flatter slope for the pure galaxy line than for the
PSF track. Since the different n/h flux ratios move in between the 
extremes of `galaxy only' and `nucleus only', they fill the range
between the two lines, flattening with increasing relative flux
contribution from the host.

At this point we can make our first quantitative statement, based on
the observed data: Small host galaxies with a half-light radius of 
only 2.4~kpc are not able to explain the extended flux we measure 
in HE\,0037--5155. However, answering the initially posed question 
whether \emph{one} host galaxy model is enough to explain the 
observed distribution is harder.

\subsection{Including noise} \label{sec:distr}

At this stage of the analysis we need to investigate how photon
shot noise translates into uncertainties in $(r_{20},r_{80})$ 
parameter space. It can be expected that noise will influence 
both parameters, but a priori the size and orientation of the 
joint error ellipses is unknown. Without this knowledge it is only
possible to give a rough estimate of the n/h flux ratio and
possible host galaxy radius. One might assume that one of the 
lines in Fig.~\ref{fig:grid} could provide a good fit, 
but for a quantitative answer we need to include random noise effects 
into the simulations.

In order to do this, artificial objects were created in the same way
as before, but these were matched to both the flux of the observed 
quasar images as well as to their noise properties.  To investigate 
the scatter expected for any constant n/h flux ratio, we selected 
five intervals in $r_{20}$ typically containing 7--8 stellar images each. 
Each of these sets shows some scatter around the PSF track which 
represents the uncertainty about the details of the PSF in the presence
of shape variations. Shot noise will be less important for the 
high S/N PSF star images. After randomly assigning one of the matching
PSF star images and adding Gaussian noise to each 
artificial object in a given $r_{20}$ slice, the artificial 
quasars contain both the background noise and the PSF shape variation.

\begin{figure}
\centering
\includegraphics[bb=189 359 410 516,clip,width=\columnwidth]{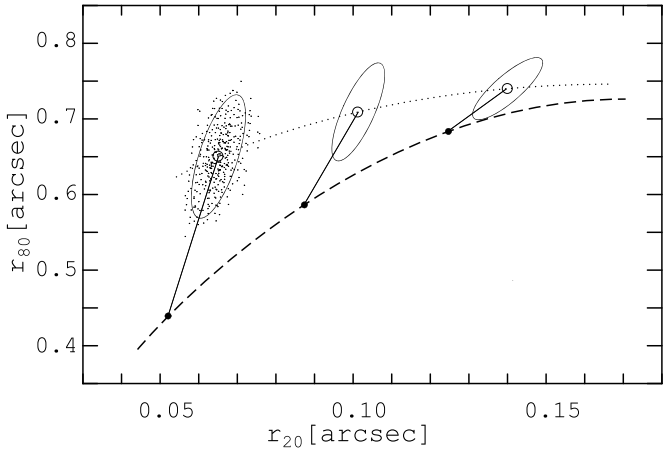}
\caption[]{Determination of error ellipses. 
Adding an artificial galaxy component to a stellar image on
the PSF track (dashed curve) shifts the location of the object
upwards and to the right, as indicated by the open circles. 
These are connected by a line of constant n/h flux ratio for 
a given quasar (dotted curve). When noise is included, the 
distribution is widened to a cloud of points. The ellipses
show the size and orientation of the derived 2$\sigma$ contours.
 \label{fig:errorell}}
\end{figure}

For each object in a slice we computed 100 different noise realizations,
leading to 100 individual pairs of $r_{20}$ and $r_{80}$.
Plotted into the $r_{20}$--$r_{80}$ diagram, the scatter can be
roughly quantified as a tilted error ellipse as shown in Fig.~\ref{fig:errorell}. 
We computed these error ellipses for both the so far assumed 
best-fitting galaxy for each object, and for the null hypothesis 
of no detectable host galaxy flux contribution. We did this for all 
three quasars individually since their S/N ratios differ significantly, 
and since we did not wish to assume a priori that the shape of the error 
ellipses were independent of S/N or flux ratio (which it however turned
out to be).

The scatter in $r_{80}$ is dominated by the shot noise in the quasar 
images, while the spread in $r_{20}$ is mainly attributable to the 
width of the stellar $r_{20}$-slices, thus in the uncertainly of
the adopted PSF. Scatter along the minor axis can therefore be reduced
if more stellar images are available in the simulations,
whereas scatter along the major axis can only be reduced by
acquiring quasar observations with higher S/N. 

An important diagnostic is the \emph{orientation} of the error ellipses 
relative to the ($r_{20}$,$r_{80}$) axes. 
Since noise will shift all points preferentially along the major
axis of the corresponding ellipse, these can now be used 
to backtrace each observed data point to the most probable
intrinsic location on a given constant n/h line from where it was scattered.
Combined with the knowledge of which -- nearly shot-noise free -- PSF
star image was used to construct the error ellipse at this location,
we can assign a unique $r_{20}$ value (that of this PSF star) to each
quasar data point. This is taken as a representation of the most
probable observing conditions under which the data was taken. We
stress that the reason for this exercise is not to try to remove noise
from an actual observation, which is obviously impossible, 
but to reveal the underlying distribution in $r_{20}$--$r_{80}$ 
space of the point sources which were folded into the quasar images.

\begin{figure}
\centering
\includegraphics[bb=184 358 407 515,clip,width=\columnwidth]{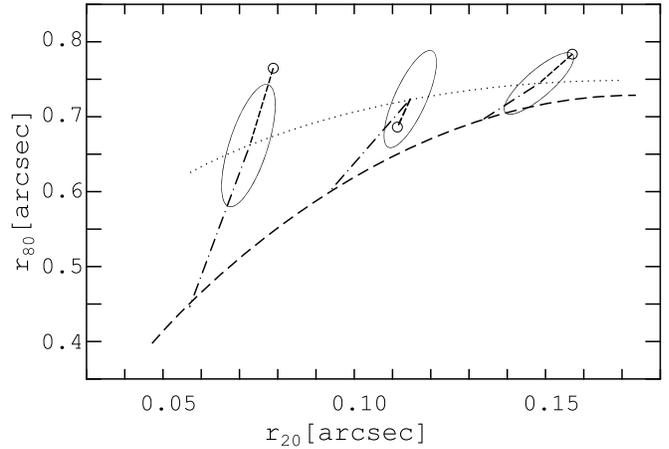}
\caption[]{Identifying the most probable PSFs for three given 
(observed) quasar images, marked by the open circles. Lines are
denoted as in Fig.~\ref{fig:errorell}. Assuming a specific n/h flux
ratio, there is one error ellipse for each quasar with the major axis 
pointing towards it. From the centre of that ellipse we can point
at the corresponding point source location on the PSF track, 
using the simulations as in Fig.~\ref{fig:errorell}. 
\label{fig:qmove}}
\end{figure}

In Fig.~\ref{fig:qmove} this task is sketched schematically. Recall
that we assume to have a best-guess constant n/h flux ratio which 
allows us to draw the corresponding curve. The first objective is 
to estimate the position on the constant n/h track from which the 
data point was most probably scattered. 
Each data point could, in principle, be derived from any point on 
the track, but there is only one error ellipse with its major axis 
directed towards the data point in question. The centre of
that ellipse represents the most probable intrinsic $r_{20}$ value.
We also know from the simulations which corresponding point source 
underlies this particular error ellipse, and we know the position 
of this point source on the PSF track. Hence, we can derive
a `reduced $r_{20}$ value' $r_{20,\mathrm{red}}$ of the quasar image.

\begin{figure}
\centering
\includegraphics[bb=50 569 288 739,clip,width=\columnwidth]{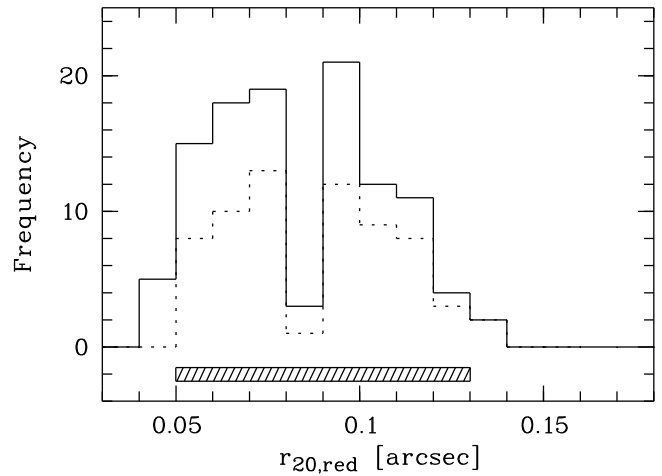}
\caption[]{
Distribution of $r_{20,\mathrm{red}}$ values for HE~0320$-$1045
(solid line), and distribution of $r_{20}$ values for the
selected matched sample of stellar images (dashed line). The
distributions agree very well in the range containing a sufficient
number of stellar images (indicated by the hashed bar). 
\label{fig:dist}}
\end{figure}

By doing so for all quasar observations, we can construct the
distribution of $r_{20,\mathrm{red}}$ values which 
essentially allows us to estimate the distribution of observing 
conditions for a given set of quasar images (cf.\ Fig.~\ref{fig:dist}).
By selecting a subsample of stellar images 
that has the same distribution in $r_{20}$ as the $r_{20,\mathrm{red}}$ 
values of a given quasar, we can now create fine-tuned simulations 
having the same flux, the same noise amplitude, \emph{and the same 
statistical PSF variation properties} as the real quasar.
We can then proceed to quantitatively test the initial hypothesis 
of assuming a certain n/h flux ratio by comparing the distribution 
of observed and simulated $r_{20}$ and $r_{80}$ values, e.g.\
by means of the standard Kolmogorov-Smirnov test. We perform 
such tests for our objects in Sect.~\ref{sec:quasars} below.

\section{Analysis of the individual quasars}  \label{sec:quasars}

The procedure outlined above provides a recipe how to
test the hypothesis that a given host galaxy model is compatible
with the observations. It does \emph{not} provide a fitting scheme, 
at least not in the strict sense. However, we can perform this
test for several model parameters and thus constrain the
range of compatible models. We also include the 
\emph{null hypothesis} that no host galaxy is detectable, and
that the observed scatter is compatible with pure noise.

Before we proceed to these tests, we briefly touch on some
preliminary considerations. Firstly, we note that the recipe to
reconstruct the $r_{20,\mathrm{red}}$ values of the quasars
relies on all quasar images to have a matching PSF star image 
to be paired with. A small number of our quasar images were
obtained under exceptionally good conditions (small $r_{20}$)
for which no matching PSF can be found. A similar though
less clear-cut situation occurs at very large $r_{20}$.
For the statistical tests we therefore deselected the extreme tails
of the $r_{20}$ distribution, using only a limited range of values
as indicated by the hashed bar in Fig.~\ref{fig:dist}. This meant
that we had to effectively throw away some of our best quasar images,
but at the gain of now having a clean matched dataset.

\begin{figure}
\centering
\includegraphics[bb=45 285 285 739,clip,width=\columnwidth]{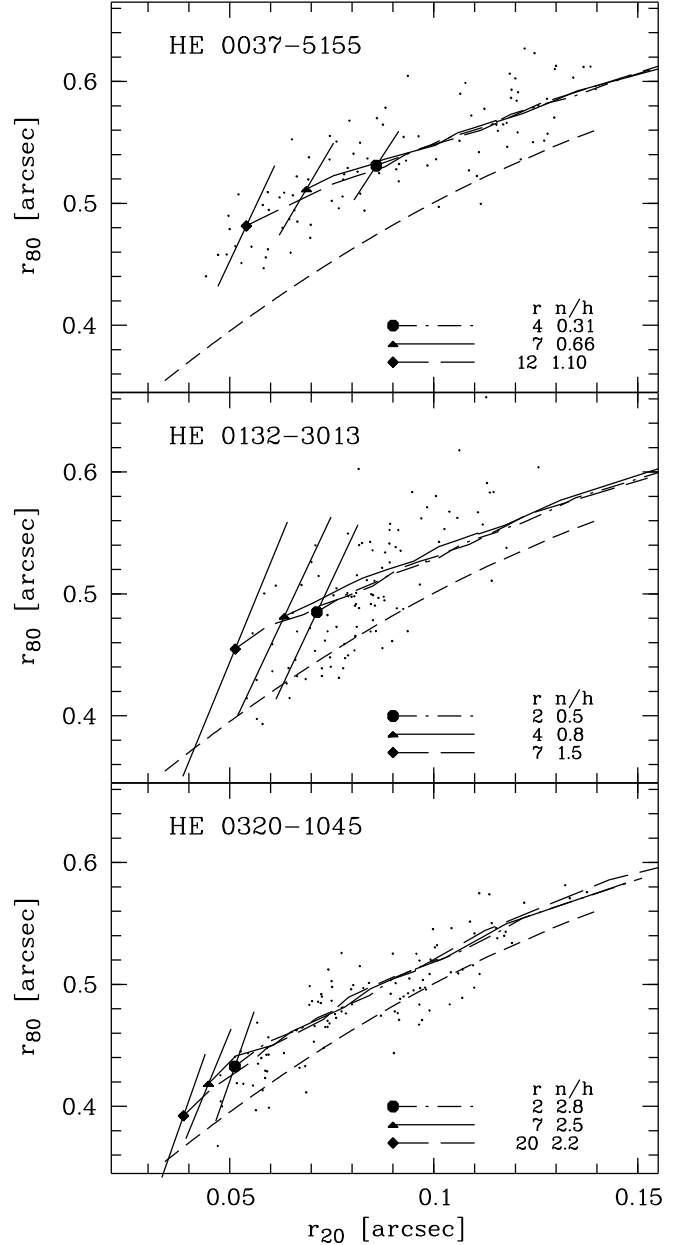}
\caption{Comparison of different models for the three quasars.
  The solid curve shows the best-fit model, the dashed and dot-dashed
  curves show two models with bracketing values of the 
  half-light radius. The beginning of each curve is marked by a filled
  symbol and a 2$\sigma$ error bar pointing along the major axis of
  the error ellipse.
  \label{fig:errest}}
\end{figure}

From Fig.~\ref{fig:grid} it is evident that our observations do
not provide strong constraints on host galaxy scale lengths.
There is a significant degeneracy between scale length and
flux ratio n/h, which is even more pronounced when noise is taken
into account. This degeneracy is illustrated in Fig.~\ref{fig:errest} 
for each of our three quasars, where we show three models with 
almost coinciding tracks. However, a closer inspection of 
Fig.~\ref{fig:errest} reveals that the degeneracy is not complete. 
Firstly, the starting point of the track for a given model 
(i.e.\ the location of the model quasar with the lowest $r_{20}$)
clearly depends on host galaxy size; a very compact host galaxy
will shift the track to the right, making it successively
less likely to reproduce the full observed distribution.
We have performed a standard Kolmogorov-Smirnov (KS) test
to compare the cumulative distributions of $r_{20}$ values
observed and predicted by different models. 
For all three quasars we can exclude the most compact of the three
models shown in Fig.~\ref{fig:errest}. In terms of a KS probability $p$, 
the 4~kpc model for HE~0037$-$5155 has $p = 0.4$~\%, 
the 2~kpc model for HE~0132$-$3013 has $p = 3.8$~\%, and the
2~kpc model for HE~0132$-$3013 has $p = 1.9$~\% probability
of acceptance. Even smaller host galaxies lead to a rapid decrease
of the probabilities. For intermediate host galaxy sizes, the 
KS probabilities are around 50~\%, and the models are fully acceptable.
Towards very large half-light radii ($\ga 12$--20~kpc) the
probabilities decrease again, but not to levels sufficiently
low for rejecting the models (this is mainly due to the lack of very 
narrow PSF star images). The one-dimensional KS test thus
provides constraints only to lower bounds to the host galaxy sizes.

However, Fig.~\ref{fig:errest} reveals that also the amplitude 
of the scatter along the major axis of the error ellipses 
varies with the size of the galaxy.
A very large galaxy will cause more scatter than a compact one,
mainly because it is more affected by shot noise, and the
distribution of scattered points will thus be widened.
This property constrains again the galaxy size, now also
discriminating against very large radii as shown in the following.

\begin{figure*}
\centering
\includegraphics[bb = 1.5cm 9.8cm 17.2cm 26.2cm,clip]{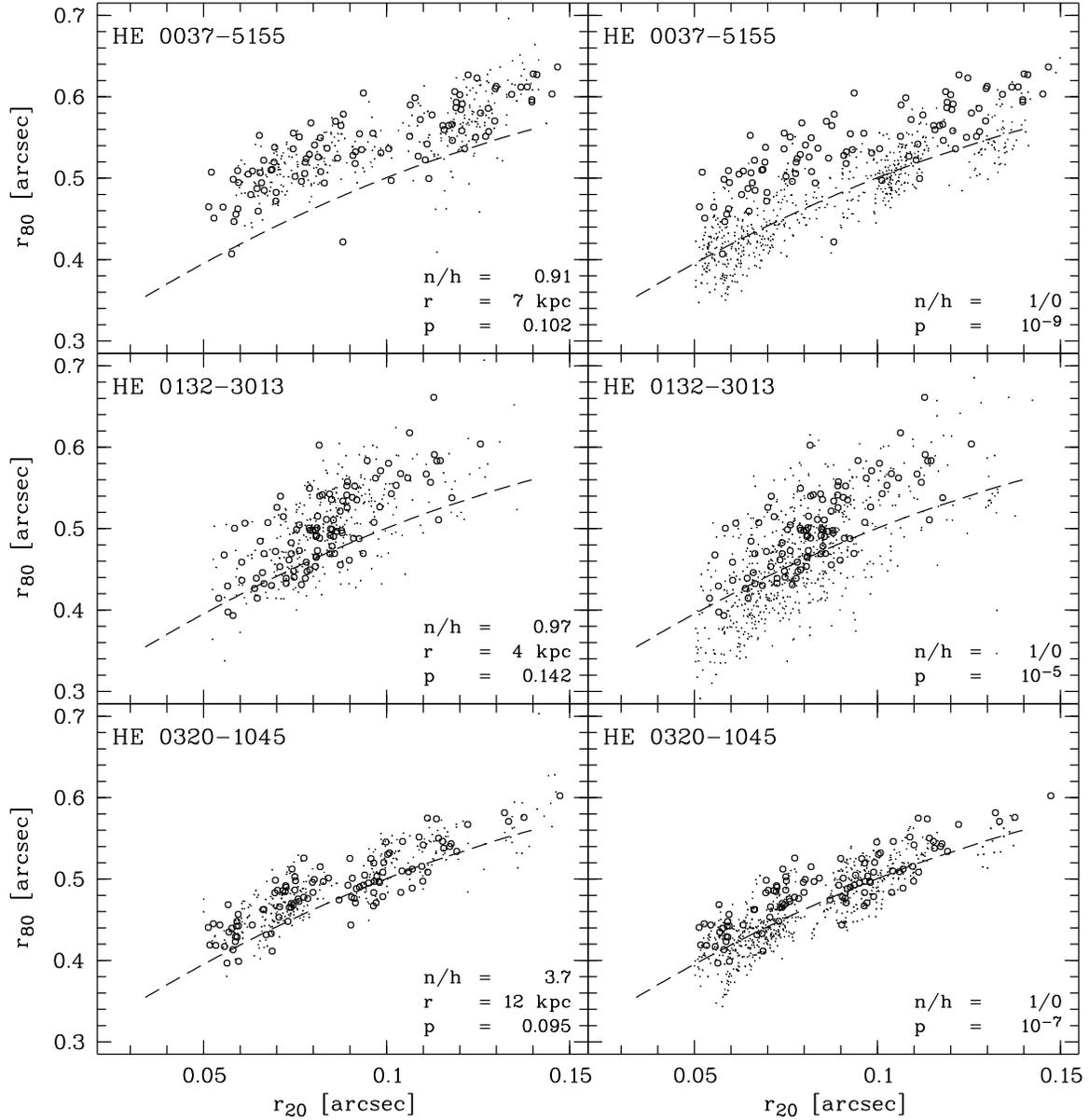}
\caption[]{Comparison of observed data and model predictions. 
  Observed quasar images are marked by circles, simulated data 
  with dots. The panels on the left show the results for our
  `best guess' models, with host galaxy fluxes and half-light radii as
  specified. For comparison, the right-hand
  panels refer to the `null hypothesis' models having
  zero host galaxy flux.
  \label{fig:fulltest}}
\end{figure*}

\begin{table*}
\centering
\caption{Two-dimensional KS test probabilities for a set of
models with different host galaxy sizes, as well as for the 
null hypothesis of zero host galaxy flux. {\bf Bold} numbers indicate the 
models with the highest probabilities.\label{tab:kserr}}
\begin{tabular}{llccccccccc}
  \noalign{\smallskip} \hline
  \noalign{\smallskip}\hline\noalign{\smallskip} 
  && \multicolumn{8}{c}{$r_{1/2}/$[kpc]}& null\\
           Object &                  &    1&    2&    4&    7&   12&   20&   35&   60&\\\hline
                  &n/h               &     & 0.25& 0.50& 0.91&1.2  & 1.2 & 1.4 & 1.4 & 1/0\\
   HE $0037-5155$ &$K^\mathrm{host}$ &     & 15.6& 15.8& 16.1& 16.2& 16.2& 16.3& 16.3& \\
                  &$p$               &     &0.004&0.015&\bf0.102&0.049&0.036&0.025&0.019&$3.6~10^{-9}$\\\hline
                  &n/h               & 0.17& 0.83& 0.97& 1.7 & 2.6 & 2.9 &3.2  & 3.2 & 1/0\\
   HE $0132-3013$ &$K^\mathrm{host}$ & 16.0& 16.5& 16.6& 16.9& 17.2& 17.4& 17.4& 17.4& \\
                  &$p$               &7~$10^{-5}$&0.008&\bf0.142& 0.084&0.036&0.028&0.008&0.001& $2.1~10^{-5}$\\\hline
                  &n/h               & 1.5 & 2.2 & 3.1 & 3.6 & 3.7 &  3.7& 3.8 & 4.0 &1/0\\
   HE $0320-1045$ &$K^\mathrm{host}$ & 15.8& 16.0& 16.3& 16.4& 16.5& 16.5& 16.5& 16.5&\\
                  &$p$               &0.004&0.007&0.022&0.033&\bf0.095&0.056&0.049&0.043& $4.1~10^{-7}$\\\hline

  \noalign{\smallskip} \hline
\end{tabular}
\end{table*}

In Fig.~\ref{fig:fulltest} we show observed and predicted
distributions of $r_{20}$ vs.\ $r_{80}$ values for our three quasars.
Each quasar is featured twice: In the right-hand panels we adopted the
null hypothesis that no host galaxy is actually detected. It is
immediately apparent that the observed and predicted distributions do
not match at all.   This is confirmed by applying the two-dimensional
KS test \citep{peacock83} which yields $p \la 0.0001$ in all three
cases. We conclude that an additional component is required in all of
our quasars, with high significance. Notice that the relevant quantity is
%
%
the distance along the displacement tracks of Fig.~\ref{fig:qmove}
(and not the proximity to the PSF tracks). Furthermore, while the
rejection of the pure PSF model is significant even in moderate seeing
conditions, any attempt to constrain the host galaxy size and hence the
n/h ratio will fail without data in the best seeing regime
(cf. Fig~\ref{fig:grid}).

We now demonstrate that assuming the presence of quasar host galaxies
with plausible physical parameters provides a statistically acceptable
explanation for these extra components. The left-hand panels of
Fig.~\ref{fig:fulltest} show the predictions of our `best guess'
models, compared again to the observed data. Notice that the selection
of PSF stars according to a matched $r_{20,\mathrm{red}}$ distribution
as illustrated in Fig.~\ref{fig:dist} ensures that even gaps in the
observed distribution (such as those apparent in the top and bottom
panels) are correctly taken into account.

There is no obvious mismatch between the observed and predicted points,
for neither of the quasars. While some of the finer details may not be
reproduced fully by the model, the overall degree of coincidence is
satisfactory. A similar conclusion is reached by looking at the 
two-dimensional KS test, which gives acceptance probabilities of
around 10--14~\% for these models (see Table~\ref{tab:kserr}). 
While this is not exactly a high probability, 
it is above the conventional threshold of 5~\% corresponding to a 
2$\sigma$ confidence level. We therefore confirm that the image
properties can be consistently described with the superposition of 
a nuclear point source plus an extended elliptical host galaxy.

Strictly speaking, this sort of consistency between model prediction
and observations is all that the KS test can provide. It is generally
not useful as a fitting tool, because of the discrete jumps of the KS
diagnostic under continuous parameter variation. We have therefore
just considered a very coarse grid of models, distinguished by
different effective host galaxy radii, and subjected these to the
two-dimensional KS test as described above. Results are also
summarised in Table~\ref{tab:kserr}.  We find again that very compact
host galaxies compare poorly with the data; in one case (HE
$0037-5155$) the offset between data points and the PSF track is so
substantial that we could not even find a `best guess' model for the
smallest host galaxy model. But now also the most extended models show
a significant decrease in their KS probabilities, due to the increased
predicted scatter as explained above.  Nevertheless we stress that the
purpose of this exercise was not to determine a 'best-fit' solution,
but just to explore the range of models that are still compatible with
the data.  Table~\ref{tab:results} summarises the results for one
representative intermediate model of each quasar (which one might call
our `best guess'), and provides relevant astrophysical parameters
connected to each model.

Because of the degeneracy between effective radius and nuclear flux
to host galaxy flux ratio n/h, the uncertainties in the size estimation
translate directly into n/h uncertainties. In Table~\ref{tab:kserr}
we provide, for each assumed galaxy size, the corresponding n/h value
and the apparent magnitude of the host galaxy.
For the two quasars with fainter nuclei HE~0037$-$5155 and HE~0132$-$3013, 
the n/h depends strongly on galaxy size, while the uncertainty  
of n/h in the very bright object HE~0320$-$1045 is much less.  
As our procedure does not generate formal 1$\sigma$ errors, we
simply adopt an uncertainty in effective radius of a factor of $\sim 2$
for the former two objects, and of a factor of $\sim 3$
for the latter, and we then estimate corresponding errors for the
host galaxy fluxes. These are also listed in Table~\ref{tab:results}.

\begin{table*}
  \begin{center}
    \caption{Results of the analysis, for our `best guess' model
      of each quasar host galaxy. n/h is 
      the flux ratio between nucleus and host; $r_{50}$ 
      is the half-light radius of the host galaxy along with the
      assumed range of possible values;
      $K_\mathrm{nuc}$ and $K_\mathrm{host}$ are the nucleus and
      host apparent $K_s$ band magnitudes at the `best guess' and most
      extreme points in the range of $r_{50}$ values; $M_R^\mathrm{nuc}$ and
      $M_R^\mathrm{host}$ are the nuclear and host absolute magnitudes in 
      the rest-frame $R$ band; $L/L_\mathrm{Edd}$ is the nuclear luminosity
      in multiples of the Eddington luminosity.}
    \begin{tabular}{lrlrrrrrrr}
\hline\noalign{\smallskip}
\hline\noalign{\smallskip}
Name & n/h & $r_{50}$ [kpc]& $K^\mathrm{nuc}$ & $K^\mathrm{host}$ & 
                 $M_R^\mathrm{nuc}$ & $M_R^\mathrm{host}$ & $L/L_\mathrm{Edd}$\\
\noalign{\smallskip}\hline\noalign{\smallskip}
HE $0037-5155$ & 0.9 & $7\pm$~0.3 dex &  $16.2^{+0.4}_{-0.1}$ & $16.1^{+0.2}_{-0.3}$ &$-27.1$ & $-27.2$ & 0.1\\\noalign{\smallskip}
HE $0132-3013$ & 1.0 & $4\pm$~0.3 dex &  $16.6^{+0.1}_{-0.3}$ & $16.6^{+0.3}_{-0.1}$ &$-26.8$ & $-26.9$ & 0.1\\\noalign{\smallskip}
HE $0320-1045$ & 3.7 & $12\pm$~0.5 dex & $15.0^ {+0.04}_{-0.01}$& $16.5^{+0.04}_{-0.1}$&$-28.5$ & $-27.0$ & 0.6\\
\noalign{\smallskip}\hline
    \end{tabular}
    \label{tab:results}
  \end{center}
\end{table*}

\begin{figure*}
\setlength{\unitlength}{1cm}
\begin{center}
\begin{picture}(17,13.2)
\put(1.4,12.6){QSO, scaled PSF and host}
\put(6.8,12.6){host galaxy}
\put(10.7,12.6){host galaxy}
\put(14.6,12.6){PSF$_1$ -- PSF$_2$}
\put(0.0,0.0){%
\includegraphics[clip]{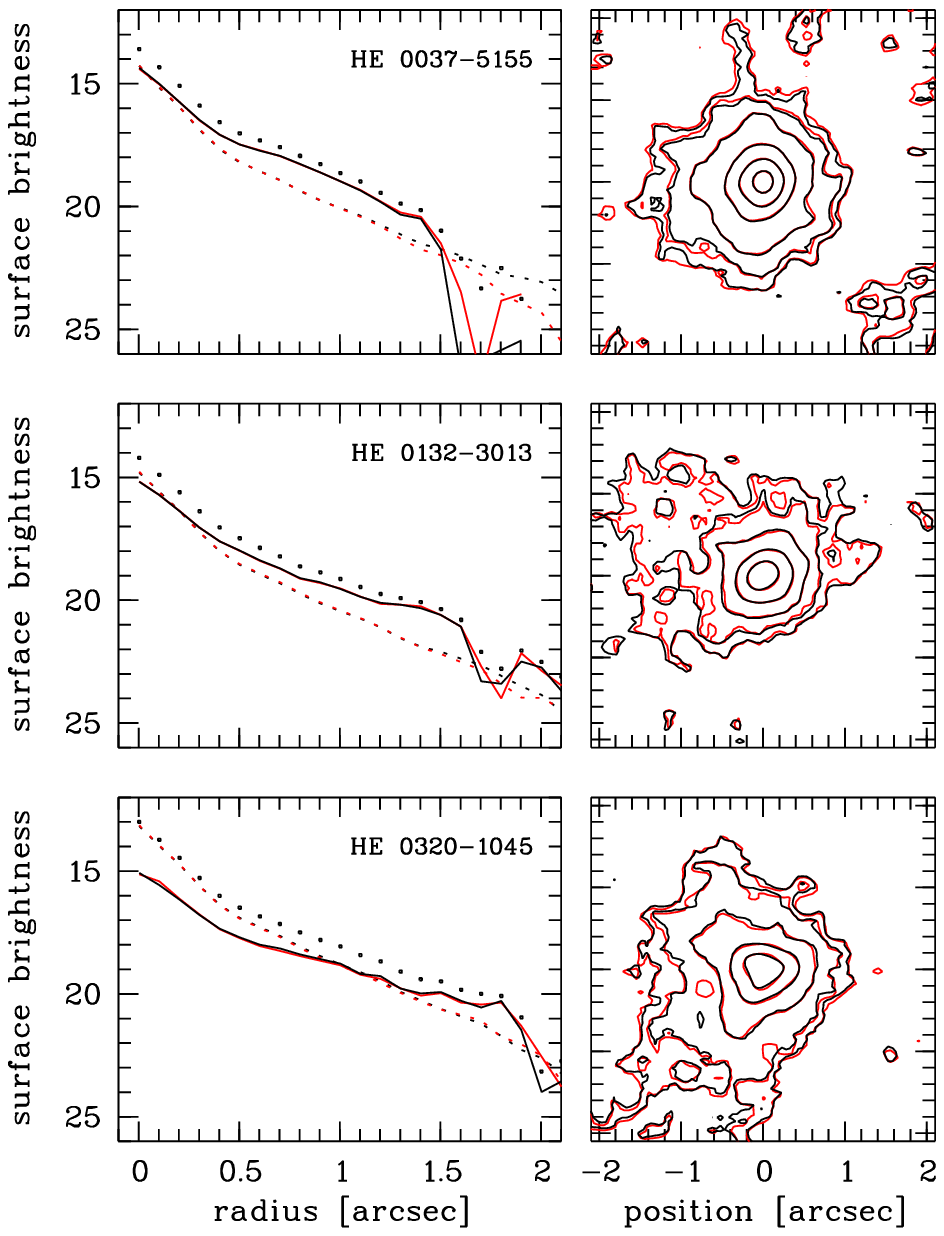}}
\put(9.8,0.9){%
\includegraphics[bb =100 100 185 185,width=3.5 cm,clip]{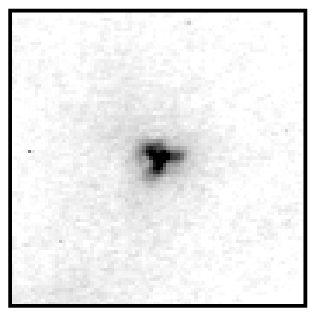}}
\put(9.8,4.9){%
\includegraphics[bb =100 100 185 185,width=3.5 cm,clip]{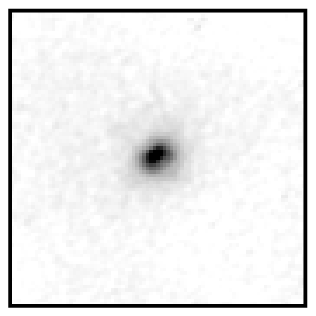}}
\put(9.8,8.9){%
\includegraphics[bb =100 100 185 185,width=3.5 cm,clip]{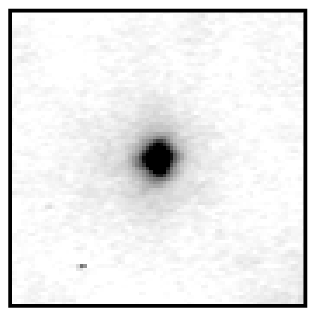}}
\put(13.9,0.9){%
\includegraphics[bb =100 100 185 185,width=3.5 cm,clip]{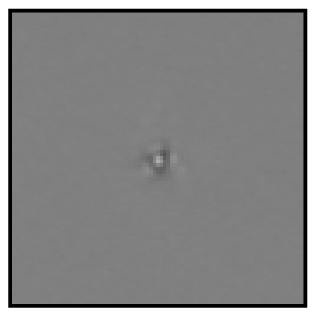}}
\put(13.9,4.9){%
\includegraphics[bb =100 100 185 185,width=3.5 cm,clip]{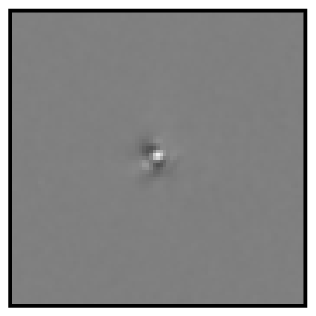}}
\put(13.9,8.9){%
\includegraphics[bb =100 100 185 185,width=3.5 cm,clip]{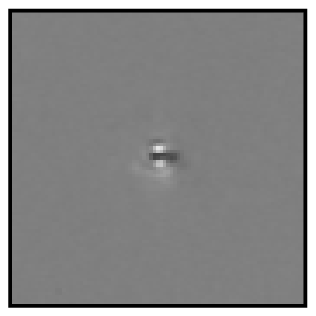}}
\end{picture}\end{center}
\caption[]{Coadded images, based on the procedure described in the text. 
  The left hand panels show azimuthally averaged surface brightness profiles
  of the quasar data (dots), the two scaled and coadded PSFs (dotted  
  lines in red and black), and the remaining host galaxy images after 
  subtraction of the different PSFs (solid lines). The middle panels show
  the quasar after PSF subtraction, both as contour and as a grayscale plot.
  The contours run at 1 mag/$\Box''$ spacing. The lowest isophote is 20
  mag/$\Box''$. The red and black contours indicate the two different PSFs used.
  The right-hand panels show grayscale images of the residuals 
  after subtracting the two PSF from each other.
  For colour versions of these figure see the online edition of the journal.
  \label{fig:coad}}
\end{figure*}

We finally computed high S/N images of each quasar by coadding only
the images with $r_{20}$ below the median, i.e.\ the 50~\% with the
best observing conditions.  To approximate a composite PSF for the
combined images, we coadded only those stellar images that had been
selected to represent the $r_{20}$ distribution of the quasar images
(e.g.\ for HE~0320$-$1045 we took the stars contributing to the dashed
histogram in Fig.~\ref{fig:dist} and inside the limits of the hashed
bar).  In order to ensure that the composition of a particular PSF had
no influence on the final PSF image, we randomly split this set of
images into two statistically independent subsets (PSF$_1$ and
PSF$_2$), i.e. each PSF image contributes to only one final PSF.
These PSF images were scaled to the quasar nuclear flux and
subtracted. In Fig.~\ref{fig:coad} we show surface brightness profiles
and contour plots of the coadded images, as well as the residuals
after subtracting the two PSF images from each other.  The two PSF
images hardly differ at all, resulting in almost indistinguishable
profiles and a residual which is confined to only the very central
parts. The total flux of the PSF residual is zero, and the scatter
around this value is less than 2~\% of the original flux per pixel.
Consequently, the two host galaxy images resulting from subtraction of
the two PSFs are almost identical, which can clearly be seen in both
the profile and the contour plots.  From this we conclude that any
structure detected in the host galaxy images is physical and not an
artefact of the PSF subtraction.

\section{Discussion}
\label{sec:disc}

We have successfully detected the host galaxies underlying all of
our three observed quasars. Resulting magnitudes and scale lengths 
are presented in Table~\ref{tab:results}. Since the
rest-frame $R$ band is virtually identical with the observed $K$ band
for all our objects, we neglected any K-corrections except 
the $(1+z)$ bandwidth term in computing the absolute magnitudes.

\begin{figure}
\centering
\includegraphics[clip,width=\columnwidth]{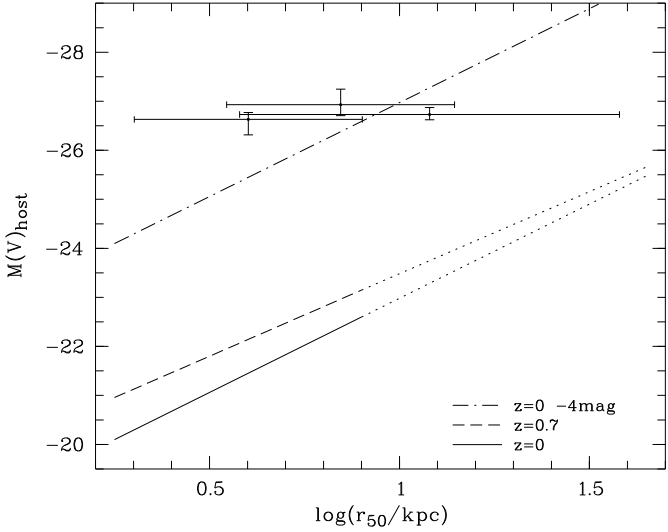}
\caption[]{
Luminosity-size relation of our three host galaxies (symbols),
compared to literature relations. For ease of comparison, the
$M_R$ magnitudes were converted to $M_V$ band assuming a rest-frame
colour $V-R=0.27\pm0.1$, appropriate for a single stellar population
of intermediate age
\citep[100~Myr to 1~Gyr, based on solar metallicity models by][]{bruz03}.
The solid line is the relation for early type galaxies at $z\simeq 0$
from \citet{shen03} in the version given by \citet{mcin05}.
The dashed line is the same relation for early type galaxies at
$z\sim0.7$ from GEMS \citep{mcin05}. The dot-dashed line is the $z=0$
relation shifted by $-$4~mag. Dotted lines mark the extrapolation
of the relations beyond the region covered by data.
\label{fig:magsize}}
\end{figure}


%
These host galaxies are intriguingly luminous, especially given their
relatively moderate sizes. In Figure 15 we relate these two quantities.
Compared with empirical $M_V$-$r_{50}$ relations of early-type
galaxies established at lower redshifts, our hosts are overluminous
by at at least $\sim 4$~mag. If our hosts are roughly ellipticals,
then they should have a much lower mass-to-light
ratio than inactive elliptical galaxies at low redshifts,
suggestive of a substantial young stellar population.
It is difficult to quantify this in the absence of colour information,
but if we simply assume that these galaxies will fade passively
to reach the luminosity-size relations at low $z$, they would need to fade
by $\sim 3.5$~mag within the next 4.2~Gyrs to $z=0.7$, or by $\sim 4$~mag
within 10.5~Gyrs to $z=0$. A single stellar population of 100--150~Myr
\citep[solar metallicity; ][]{bruz03} would have this property, fading
by $\sim$3.2~mag and $\sim$4.0~mag in 4.2~Gyr and 10.5~Gyrs,
respectively. Although this involves a lot of assumptions, the
conclusion that we have detected the signature of a significant
young stellar population is consistent with recent results
from GEMS \citep{jahn04} where similar ages were found for
AGN hosts at $1.8<z<2.5$.

\begin{figure}
\centering
\includegraphics[bb = 55 451 399 774,width=\columnwidth,clip]{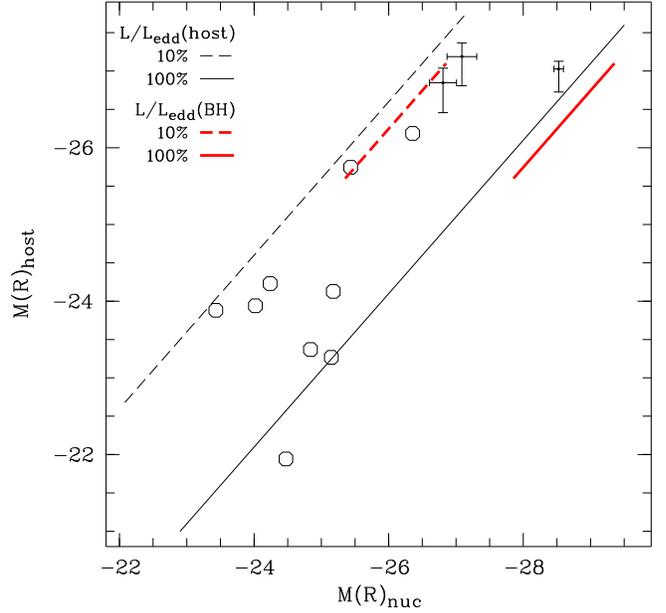}
\caption[]{Nuclear versus host luminosities. Dots with error bars mark our
  data, circles mark hosts from the $z\sim2$ sample of
  \protect{\citet{kuku01}}. We also plot lines of constant
  Eddington ratio derived from the host galaxy luminosities,
  extrapolated from low $z$ by assuming only passive evolution 
  (thin black lines). The short thick-lines segments denote 
  Eddington ratios derived from the spectroscopic black hole 
  masses of our objects.
  \label{fig:nuchost}}
\end{figure}

In Fig.~\ref{fig:nuchost} we plot nuclear luminosities against host
luminosities for our objects. We also include the $z\sim 2$ results of
\citet{kuku01}, transformed to the rest-frame $R$ band assuming
$(V-R)_{nuc}=0.4$, and zeropointing to the Vega system. This type of
diagram has very often been constructed for quasar host galaxies at
lower redshifts, showing clearly a correlation between quasar nuclear
and host galaxy absolute magnitudes \citep{mcle95}.
\citet{mcle99}
demonstrated that this correlation is due to a physically motivated
diagonal in this diagram, below and right of which there are no
quasars because they would exceed their respective Eddington limits.
In drawing this line, one has to assume a relation between host galaxy
(spheroid or bulge) luminosity and the mass of the central black hole.
While at low redshifts such a relation now seems well established
\citep{mclu02,marc03,haering04}, it is by no means clear how a
corresponding relation at $z \sim 2$ should look like. As a naive
first guess we assumed that the same relation holds as at low
redshifts, taking into account only passive evolution for 
 a formation epoch at $z = 3$. 
This is shown by the thin lines in
Fig.~\ref{fig:nuchost} which mark the loci of quasars radiating at
10~\% and 100~\% of their Eddington luminosities.
  
Our quasars have nuclear absolute magnitudes that are on average 
$\sim 3$~mag brighter than those of \citet{kuku01}. If the above
correlation between nuclear and host luminosities pertains at these
redshifts, our host galaxy luminosities should be brighter by a 
similar amount. This is indeed the case; our three datapoints
form a seamless continuation of the \citet{kuku01} results,
roughly along a diagonal line through the diagram. This implies
that while our quasars have indeed very luminous host galaxies,
they are not excessively luminous and fit well into the context
of existing data.

So far, this does not mean 
 that the \citet{mcle99} relation still holds at high redshifts, 
as the black
hole masses in our quasars are essentially unknown and need not
be coupled to the host galaxy luminosities as assumed above.
However, recent work has opened an avenue to estimate, at least
roughly, black hole masses using the width of the
C$\:${\sc iv} emission line together with the continuum flux. The
underlying assumption is that the motion of the line-emitting gas is
virialized and that continuum flux and emission line width are
estimators for the radius and the Keplerian velocity of the broad line
region.  A relation between $M_\mathrm {BH}$, the H$\beta$ line
width and the optical continuum flux is established
\citep[e.g.][]{wand99,kasp00} and has recently been extended to
Mg$\:${\sc ii} \citep{mclu02b} and C$\:${\sc iv} \citep{vest02}.
We have used the formula from the latter paper,
\begin{equation}
  \label{eq:vester}
  M_\mathrm{BH} = 1.6\times 10^6 
  \left[\frac{\mathrm{FWHM(C\:{\mbox{\sc iv}})}}{1000 \mathrm{\:km\:s^{-1}}}\right]^2
  \left[\frac{\lambda L_\lambda(1350\:\mathrm{\AA})}{10^{44}\mathrm{\:ergs\:s^{-1}}}\right]^{0.7}
\end{equation}
to predict virial black hole mass estimates for our quasars, using
the spectra shown in Fig.~\ref{fig:spectra}.
Results are summarised in Table~\ref{tab:bhmasses}.
The approach has certainly large intrinsic uncertainties 
due to the unknown individual geometry and radiation anisotropy.
Individual masses are probably not better determined than to $\pm 0.5$~dex,
but unless the method is heavily biased, sample averages can be
quite useful. For our small sample, we obtain a mean black hole mass
of $3\times 10^{9}\:M_\odot$, with an uncertainty of at least 0.3~dex.
These masses are consistent with predictions from low- to intermediate-redshift
results \citep{mclu02} as well as with actual high-redshift data \citep{mclu03}.

\begin{table}[bt]
\centering
\caption{Emission line widths of \ion{C}{IV} and continuum fluxes at 
  1350~\AA , together with the estimated black hole masses.
  \label{tab:bhmasses}}
\begin{tabular}{lrcc}
  \noalign{\smallskip} \hline
  \noalign{\smallskip}\hline\noalign{\smallskip} 
  Object & ${\scriptstyle \mathrm{FWHM} \atop\scriptstyle\mathrm{\:km\:s^{-1}}}$ 
         &$\log\left(\frac{\scriptstyle\lambda L_\lambda}{\scriptstyle\mathrm{\:ergs\:s^{-1}}}\right)$
         & $\log\left(\frac{\scriptstyle M_\mathrm{BH}}{\scriptstyle M_\odot}\right)$\\
  \noalign{\smallskip}\hline\noalign{\smallskip} 
  HE $0037-5155$ & 5800 & 46.7 & 9.7 \\ 
  HE $0108-5952$ & 6000 & 46.3 & 9.4 \\ 
  HE $0132-3013$ & 5000 & 46.6 & 9.5 \\ 
  HE $0320-1045$ & 3300 & 47.1 & 9.4 \\ 
  HE $0418-0619$ & 6200 & 46.3 & 9.4 \\ 
  \noalign{\smallskip} 
\hline
\end{tabular}
\end{table}

We now can compare each quasar absolute magnitude to its black hole mass
and derive the Eddington ratio $L_{\mathrm{nuc}}/L_{\mathrm{Edd}}$,
assuming mean quasar bolometric corrections taken from \citet{elvi94}.
Comparing these Eddington ratios with those obtained from the host galaxy
luminosities we find that these are in very good agreement.
For illustration, we have used the spectroscopically constrained
Eddington ratios to predict where a quasar radiating at 100~\% and
at 10~\% Eddington would be located in Fig.~\ref{fig:nuchost} 
(short thick lines). These line segments are displaced by less
than a factor of 2 relative to the thin lines based on extrapolating
the low-redshift relation, well within the uncertainties of
our estimates. At least for our small set of very high luminosity
quasars, these results are consistent with the notion that their 
host galaxies and black holes follow essentially the same relation
as their present-day equivalents. 

There is, however, a caveat to this statement. It implicitly assumes 
that our host luminosities are entirely due to stellar light
from a homogeneous, well behaved stellar population. We now
consider some possibilities that could devalidate this assumption.

A first effect could be the contamination by emission lines. While
this is usually avoided by judiciously selecting the quasar sample
such that no emission lines fall inside the filter regions, 
the very tight requirements of the AO system made it impossible
for us to restrict the sample to redshift bins free of emission 
lines. In particular, H$\alpha$ lies perfectly inside the filter 
profile for all of our objects. Any extended H$\alpha$ contribution
would lead to a corresponding reduction of the stellar light inside
the $K$ band. Basically, there are three possible origins for 
extended H$\alpha$ emission: A large extended emission line region
(EELR) powered by the quasar; H$\:${\sc ii} regions excited by
young stars; or scattered light from the quasar nucleus.
The last option can probably be dismissed, since the scattered
H$\alpha$ flux in radio galaxies at $z \sim 1$ 
has been shown to be $\le 10$~\% of the nuclear flux
\citep{leys98,rigl92}.  Assuming a scattered nuclear light
fraction of this order results in a negligible contribution to the
host galaxy magnitude for the objects having n/h~$<1$. In the case of
HE~0320$-$1045 the strength of the nuclear emission could contribute
to the extended flux, though the actual amount of scattered flux
cannot be quantified without access to colour information.

\citet{moor00} investigated a sample of $z\sim 2.2$ 
H$\alpha$-emitting galaxies and found that 
$L_{\mathrm{H}\alpha}/L_K\approx 1/5$ where $L_K$ is the
luminosity of the galaxy in the $K$ band.  If these results 
are portable to our quasar host galaxies \citep[which is
not unreasonable, see][]{vilc98}, line emission from star-forming
regions may increase the host galaxy luminosity by up to 0.2
magnitudes. A similar contribution of up to 0.1~magnitudes of
H$\alpha$ to the $R$ band flux was found for low redshift quasar host
galaxies \citep{jahn03}, in this case including possible EELR
contributions. 

Another significant contamination to the host galaxy luminosity could
come from close companions. In other studies of host galaxies at high
redshift this is a relatively common feature,  and $\sim 40$~\% of the
objects analysed by \citet{lehn99}, \citet{ridg01}, \citet{hutc95},
\citet{sanc04}, and \citet{jahn04} show companions.  Since the
effective field size used here is only $6\farcs 4\times 6\farcs 4$ due
to the chopping between quadrants, no conclusions can be drawn on the
density of field galaxies in the vicinity of our objects. However, in
the direct images (Fig.~\ref{fig:coad}) we see no signs of foreground
galaxies or companions for two of the quasars, and their host galaxies
appear quite round and undisturbed.  HE~0320$-$1045, on the other
hand, shows two extended features at $\sim 10$~kpc separation to the
NE and SE of the nucleus which contain $\sim 50$~\% of the host galaxy
flux inside an aperture of 0\farcs6.  In combination with the very
pronounced core in the host galaxy luminosity profile (solid line in
Fig.~\ref{fig:coad}), this clearly indicates a host galaxy which is
disturbed, accompanied by other galaxies or in the process of merging.

We conclude that the cumulative effect of all these mechanisms 
cannot be large, probably below 0.5~mag. The origin of the
high luminosities of our quasar host galaxies must therefore 
be light from stars.
We have shown that there is evidence for a relatively low
mass-to-light ratio in these systems, and 
most of the rest-frame optical light in these quasar hosts 
could originate in young stars.
Without colour information, however, we have no handle on 
estimating even rough stellar population properties.
With\break 4--12~kpc, the sizes of these objects fall within
the range of 3--20~kpc found for less luminous quasar hosts 
at only slightly lower redshifts by \citet{ridg01}, \citet{falo01} and
\citet{falo04}.  
Given their sizes, these galaxies are
clearly much more luminous than elliptical galaxies and QSO hosts at
$z\simeq 0$ following the \citet{kormendy77} relation
\citep[e.g.][]{mclu99}. 
Further and more accurate measurements will be needed to 
quantitatively constrain the colours and sizes of such systems 
at high redshifts.

\section{Conclusions}

We have detected the host galaxies underlying three highly
luminous $z\sim 2.2$ quasars, using near-infrared adaptive optics
observations. We could estimate nuclear and host galaxy luminosities
as well as constrain their sizes, but there is some degeneracy
between these parameters. The measured host luminosities 
are among the highest yet measured at these redshifts, but
the quasars are also among the most luminous to be resolved.
Our measurements are in good agreement with other results
obtained for similar redshifts, and the location of the
host galaxies in a $L_{\mathrm{nuc}}$ vs.\ $L_{\mathrm{host}}$ 
diagram suggests that the quasars radiate at roughly 10--50~\%
of their Eddington luminosities, similar to low-redshift quasars.
Comparing our results to the magnitude-size relation at $z=0$ and $z=0.7$
indicates a much lower mass-to-light ratio in the host galaxies than
for an old population. A dominating stellar population of
100--150~Myrs age would have a passive fading to allow an evolution of
our host galaxies onto the lower redshift magnitude-size relation.

In getting these results, we have both taken advantage of, and
suffered from the special conditions occurring when working with
adaptive optics. The capability of measuring the scale length of a
$z\sim 2$ galaxy in the presence of a bright nucleus is certainly
owned to the high angular resolution achieved, especially as
our quasar hosts appear to be rather compact.
On the other hand, we had to overcome several difficulties 
caused by the strong dependence of the image quality on the 
actual atmospheric conditions.

Careful determination of the PSF is a condition without exception
for the analysis of host galaxies of luminous quasars. For AO observations 
this usually poses a big problem, because of the generally small
field of view and the strong field anisoplanatism. We have described
a conceptually simple procedure that enabled us to incorporate 
non-simultaneously observed PSF stars in a straightforward manner, 
even though the image quality varied substantially during the
observing sessions. 

While the performance of current AO instruments has certainly 
improved dramatically compared to the relatively ancient system
that we used, some of the problems mentioned above are still
the same. Detectors have become larger, but since larger telescopes
require smaller pixels in order to sample the PSF core adequately,
the field of view has not necessarily grown in proportion. 
This means that also in the future, quasars with a suitable PSF star 
in the same field of view \emph{and} the same degree of field
anisoplanatism will be very rare; thus, non-simultaneous PSF
observations will often be necessary also in the future.
One of the objectives of this paper is to create awareness
that non-optimal atmospheric conditions do exist, and that
one needs to account for rapid variability of observing 
conditions on all time scales. In this sense, AO observations
are much more difficult to handle than normal seeing-limited
imaging. Indeed, near-infrared imaging under good seeing conditions
is probably at least competitive to AO as long as mere detections
of high-redshift quasar hosts are sought. On the other hand, 
AO data provide insights into quasar host structural properties 
on kpc scales. In this domain, AO observations will be unbeatable from
the ground, and further development of specific analysis tools will be
well worth the effort.

\begin{acknowledgements}

  This work was supported by the DFG under grants Wi~1369/5--1 and
  Re~353/45--3.  E\"O acknowledges travel support from The Swedish
  Institute and The Royal Swedish Academy of Sciences.  KJ
  acknowledges support by the `Studienstiftung des deutschen Volkes'.
  This study is based on observations made with ESO Telescopes at the
  La Silla Observatory under programme ID 64.P-0411. Use was also made
  of NASA's Astrophysics Data System Abstract Service.

\end{acknowledgements}

\bibliographystyle{aa} 
\bibliography{ao}  

\end{document}